\documentclass{iaus}
\usepackage{graphicx,subfigure}

\title[Quenching SF in NGC~1266] 
{Quenching of Star Formation in Molecular Outflow Host NGC~1266}

\author[K. Alatalo and the ATLAS$^{\rm 3D}$ team]   
{K.\ Alatalo$^1$, K.\ E.\ Nyland$^2$, G.\ Graves$^1$, S.\ Deustua$^3$, L.~M. Young$^2$, T.~A. Davis$^4$, M. Bureau$^5$, E. Bayet,$^5$ L.\ Blitz,$^2$ M.\ Bois,$^6$ F.\ Bournaud,$^7$, M. Cappellari,$^5$ R.\ L.\ Davies,$^5$ P.~T.\ de~Zeeuw,$^{4,12}$ E.\ Emsellem,$^{4,15}$ S.\ Khochfar,$^8$ D.\ Krajnovic,$^4$ H.\ Kuntschner,$^4$ R.\ M.\ McDermid,$^{9}$ R.\ Morganti,$^{10}$ T. Naab,$^{8}$ T. Oosterloo,$^{10}$ M.\ Sarzi,$^{11}$ N. Scott,$^{13}$ P.\ Serra$^{11}$ and A.\ Weijmans$^{14}$}

\affiliation{
$^1$University of California, Berkeley, USA;
$^2$New Mexico Tech, Socorro, USA;
$^3$Space Telescope Science Institute, Baltimore, USA;
$^4$ESO, Garching, Germany;
$^5$University of Oxford, UK;
$^6$Observatoire de Paris, France;
$^7$Universit\'e Paris Diderot, France;
$^8$MPI for Extraterrestrial Physics, Garching, Germany;
$^9$Gemini Observatory, Hilo, USA;
$^{10}$ASTRON, Dwingeloo, The Netherlands;
$^{11}$University of Hertfordshire, Hatfield, UK;
$^{12}$Leiden University, The Netherlands;
$^{13}$Swinburne University, Australia;
$^{14}$University of Toronto, Canada;
$^{15}$Universit\'{e} de Lyon, France}

\pubyear{2012}
\volume{292}  
\pagerange{119--126}
\jname{IAU Symposium 292.~~Molecular Gas, Dust and Star Formation in Galaxies}
\editors{Tony Wong \& Juergen Ott, eds.}
\begin{document}

\maketitle
\firstsection 

\noindent{\bf Summary:}
We detail the rich molecular story of NGC~1266, its serendipitous discovery within the ATLAS$^{\rm 3D}$ survey (Cappellari et al. 2011) and how it plays host to an AGN-driven molecular outflow, potentially quenching all of its star formation (SF) within the next \hbox{100 Myr}. While major mergers appear to play a role in instigating outflows in other systems, deep imaging of NGC 1266 as well as stellar kinematic observations from {\tt SAURON}, have failed to provide evidence that NGC 1266 has recently been involved in a major interaction. The molecular gas and the instantaneous SF tracers indicate that the current sites of star formation are located in a hypercompact disk within 200 pc of the nucleus (Fig. 1; SF rate $\approx 2~M_\odot$ yr$^{-1}$). On the other hand, tracers of {\em recent} star formation, such as the H$\beta$ absorption map from {\tt SAURON} and stellar population analysis show that the young stars are distributed throughout a larger area of the galaxy than current star formation. As the AGN at the center of NGC 1266 continues to drive cold gas out of the galaxy, we expect star formation rates to decline as the star formation is ultimately quenched.  Thus, NGC 1266 is in the midst of a key portion of its evolution and continued studies of this unique galaxy may help improve our understanding of how galaxies transition from the blue to the red sequence (Alatalo et al. 2011).

\begin{figure}[h]
\vspace*{-1 mm}
\subfigure{ \includegraphics[height=1.3in]{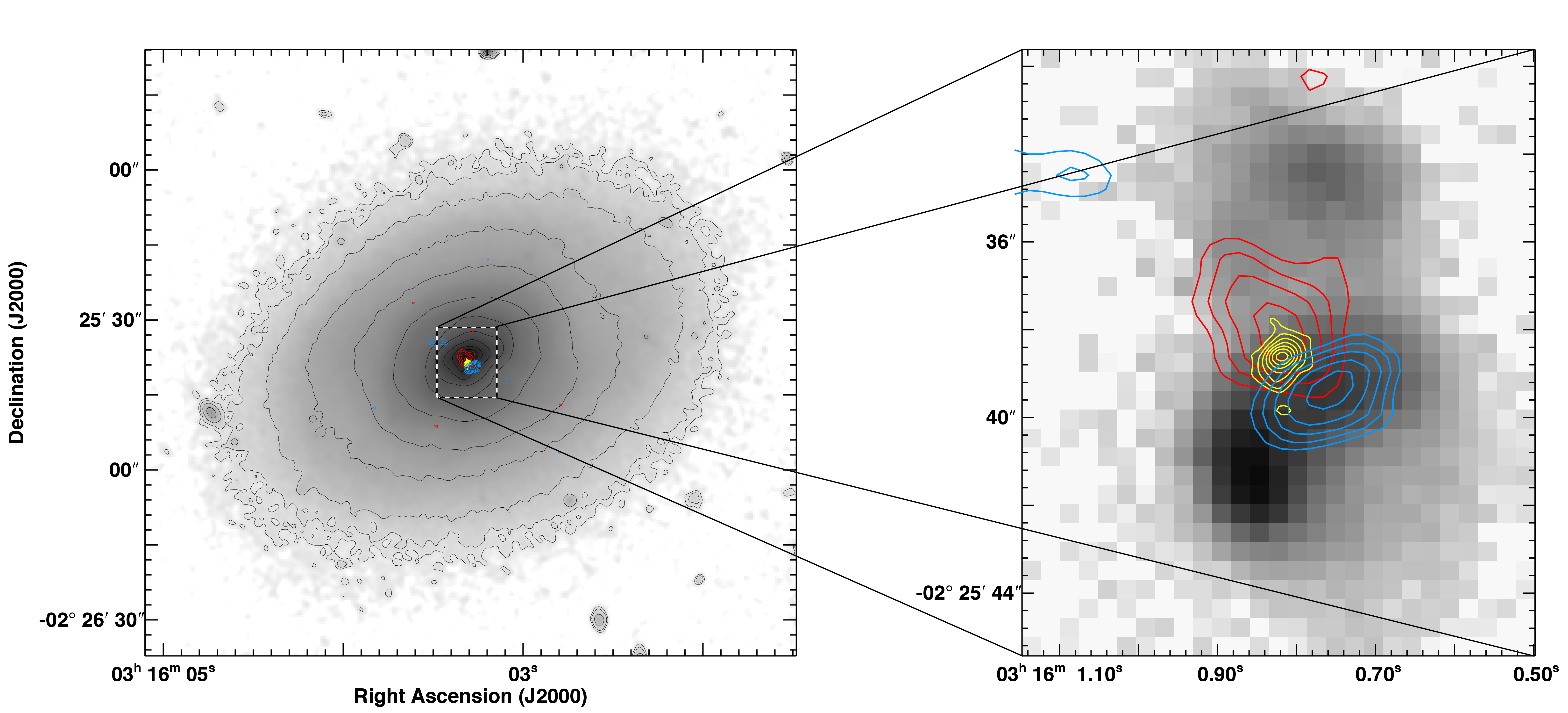} } \hspace{2in}
\subfigure{ \includegraphics[height=1.3in]{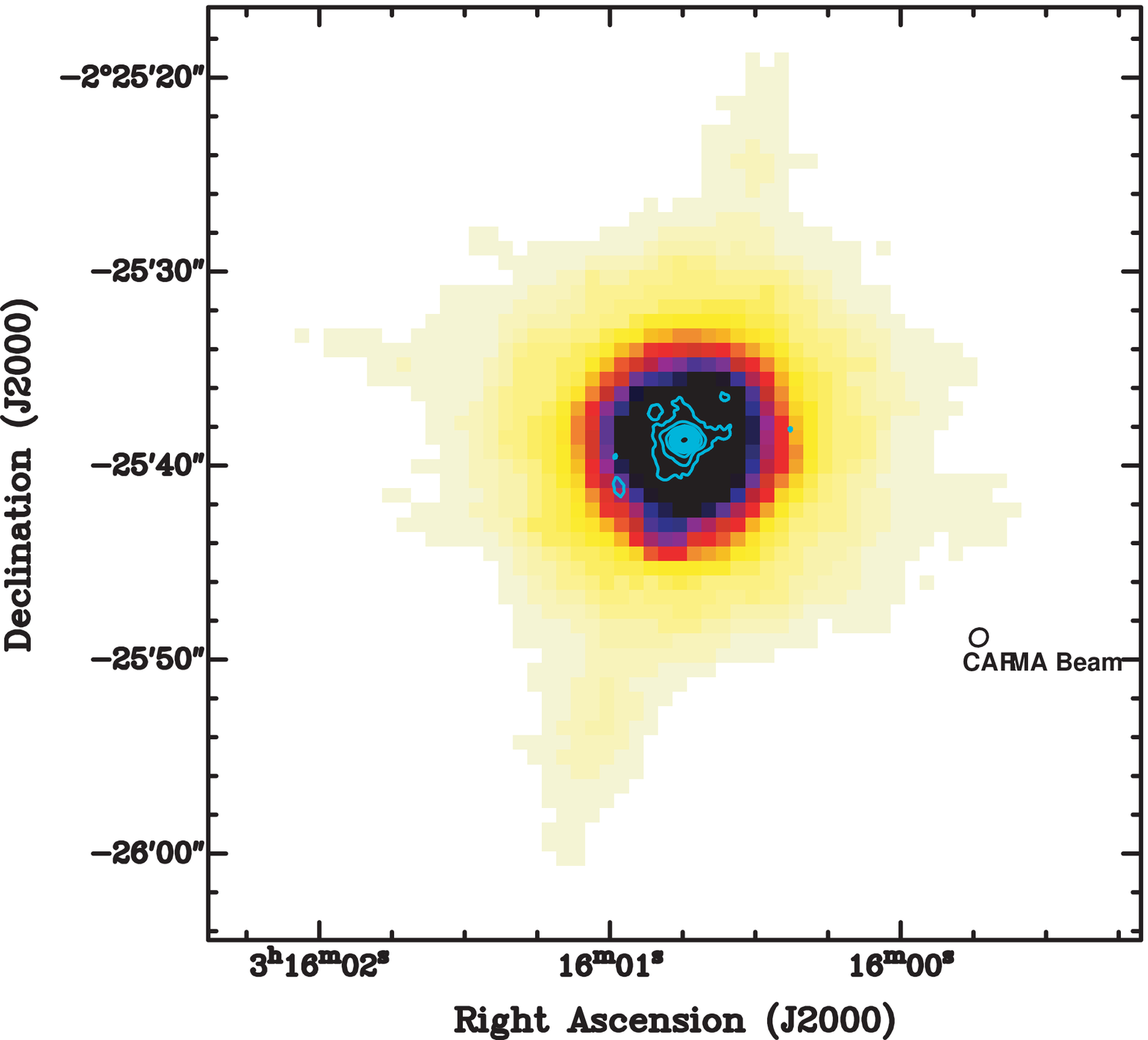} }
\vspace*{-2 mm}
 \caption{{\bf (Left):} The molecular outflow of NGC~1266 is shown on an $R$-band image, showing that its detected extent is quite small.  The zoomed-in image shows the same molecular gas overlaid on H$\alpha$ emission.  {\bf (Right):} The central molecular gas is overlaid on the {\em Spitzer} 8$\mu$m PAH emission.}
   \label{fig1}
\end{figure}


\vspace{-8mm}
\begin{thebibliography}{}
\vspace{-2mm}
\bibitem[Alatalo \etal\ (2011)]{alatalo+11}Alatalo, K., et al. \ 2011, ApJ, 735, 88
\bibitem[\protect\citeauthoryear{Cappellari et al.}{2011}]{cappellari+11} Cappellari, M. et al. \ 2011, MNRAS,  413, 813
\end{thebibliography}
\end{document}